\pdfoutput=1

\documentclass[11pt]{article}

\usepackage[]{acl}

\usepackage{times}
\usepackage{latexsym}
\usepackage{booktabs}
\usepackage{xcolor}
\usepackage{multirow}
\usepackage{graphicx}

\usepackage[T1]{fontenc}

\usepackage[utf8]{inputenc}

\usepackage{microtype}

\usepackage{amssymb}
\usepackage{subcaption}
\title{XLAVS-R: Cross-Lingual Audio-Visual Speech Representation Learning \\ for Noise-Robust Speech Perception}

\author{HyoJung Han$^A$\thanks{\quad Work done at Meta AI}\quad Mohamed Anwar\footnotemark[1]\quad Juan Pino$^V$\quad Wei-Ning Hsu$^V$ \\
\textbf{Marine Carpuat}$^A$\quad\textbf{Bowen Shi}$^V$\quad\textbf{Changhan Wang}$^V$\\
  $^A$University of Maryland, USA\quad$^V$Meta AI, USA \\
  \texttt{hjhan@cs.umd.edu}\quad \texttt{changhan@meta.com} \\
  }

\begin{document}
\maketitle
\begin{abstract}
Speech recognition and translation systems perform poorly on noisy inputs, which are frequent in realistic environments. Augmenting these systems with visual signals has the potential to improve robustness to noise.  
However, audio-visual (AV) data is only available in limited amounts and for fewer languages than audio-only resources.
To address this gap, we present XLAVS-R, a cross-lingual audio-visual speech representation model for noise-robust speech recognition and translation in over 100 languages. It is designed to maximize the benefits of limited multilingual AV pre-training data, by building on top of audio-only multilingual pre-training and simplifying existing pre-training schemes. 
Extensive evaluation on the MuAViC benchmark shows the strength of XLAVS-R on downstream audio-visual speech recognition and translation tasks, where it outperforms the previous state of the art by up to 18.5\% WER and 4.7 BLEU given noisy AV inputs, and enables strong zero-shot audio-visual ability with audio-only fine-tuning.

\end{abstract}

\section{Introduction}
\label{sec:intro}

Speech recognition and speech-to-text translation, two core speech perception tasks, have witnessed rapid developments in the past two years. 
Still, the performance of state-of-the-art (SOTA) models such as Whisper~\cite{radford2022robust} and SeamlessM4T~\cite{communication2023seamlessm4t} degrades sharply in noisy environments~\cite{anwar2023muavic,communication2023seamless}. Audio-only approaches to make such systems more robust \citep{Ng2023DehubertDN, zhu2023robust} remain challenged by frequent noise types such as intense babble noise and overlapped speech~\cite{shi22_interspeech}.

\begin{table}[t]
    \centering
    \fontsize{7}{10}\selectfont
    \begin{tabular}{lcccc}
    \toprule
    & \multicolumn{2}{c}{Hours} & \multicolumn{2}{c}{Languages} \\
    \cmidrule(lr){2-3}\cmidrule(lr){4-5}
    & A & AV & A & AV \\
    \midrule[\heavyrulewidth]
    \multicolumn{5}{l}{\textit{Audio-Only Pre-Training Data}} \\
    AVFormer~\cite{seo2023avformer} & 60K & 0 & 1$^\star$ & 0 \\
    FAVA~\cite{may2023audiovisual} & 2.8K & 0 & $>$100 & 0 \\ 
    \midrule
    \multicolumn{5}{l}{\textit{Audio-Visual Pre-Training Data}} \\
    AV-HuBERT~\cite{shi2022learning} & 0 & 1.8K & 0 & 1$^\star$ \\
    AV-data2vec~\cite{lian2023av} & 0 & 1.8K & 0 & 1$^\star$ \\
    AV2AV~\cite{choi2023av2av} & 0 & 7K & 0 & $>$100 \\
    \midrule
    \multicolumn{5}{l}{\textit{Audio-Only \& Audio-Visual Pre-Training Data}} \\
    u-HuBERT~\cite{hsu2022uhubert} & 0.5K & 1.8K & 1$^\star$ & 1$^\star$ \\
    VATLM~\cite{zhu2023vatlm} & 4.3K & 1.8K & 1$^\star$ & 1$^\star$ \\
    \textbf{XLAVS-R} (this work) & 436K & 1.2K & 128 & 9 \\
    \textbf{XLAVS-R} (+ extended AV data) & 436K & 8.3K & 128 & 100+ \\
    \bottomrule
    \end{tabular}
    \caption{Pre-training data type, amount and language coverage in audio-visual speech perception models (A: audio, AV: audio-visual). Compared to prior work, XLAVS-R exploits audio-only speech data for efficient data scaling and language coverage expansion. 
    $^\star$ English-only. 
    }
    \label{table:data}
\end{table}

Complementing audio inputs with visual signals such as video frames of speaker's lips offers an alternative approach \citep{sumby1954visual,potamianos2004audio,mroueh2015deep,son2017lip}, with promising results in audio-visual speech recognition (\textbf{AVSR}) and audio-visual speech-to-text translation (\textbf{AVS2TT}) tasks~\cite{Afouras2018DeepAS,ma2021end,shi22_interspeech,anwar2023muavic}. 
Yet, as summarized in Table~\ref{table:data}, prior works are either restricted to English \citep{hsu2022uhubert, zhu2023vatlm, seo2023avformer} or to a limited amount of audio-visual data \cite{shi2022learning,choi2023av2av,lian2023av}, limiting language coverage and performance on downstream tasks. 

To bridge this gap, we present \textbf{XLAVS-R}, a cross-lingual audio-visual speech representation learning model for noise-robust speech perception across over 100 languages. Its training strategy is designed to maximize the efficiency of multilingual AV pre-training by injecting audio-visual (\textbf{AV}) signals (\textsection \ref{sec:xlavsr_visual_modality_injection}) after pre-training on audio-only (\textbf{A-only}) data that is easier to scale in size and language coverage (\textsection \ref{sec:xlavsr_audioonly_speech_rep_learning}).
We improve AV-HuBERT~\citep{shi2022learning} with simplified training protocol and updated model architecture, and scale model size up to 2B parameters for performance in the multilingual setting.

We conduct extensive evaluations on the MuAViC benchmark~\citep{anwar2023muavic}. XLAVS-R yields SOTA performance on speech recognition in 9 languages and speech-to-text translation in 6 X-to-English pairs (\textsection \ref{sec:sr_results}, \textsection\ref{sec:st_results}), notably improving their robustness to noisy inputs. Furthermore, XLAVS-R enables zero-shot audio-visual ability in downstream tasks with audio-only fine-tuning (\textbf{FT}): fine-tuning our XLAVS-R 2B model without supervised AV downstream data achieves the best audio-visual performances in noisy settings even compared to the corresponding audio-visual supervised fine-tuned model (\textsection \ref{sec:results_zero_shot}). This confirms the benefits of our pre-training strategy and relaxes the data requirements for downstream tasks, making it possible to tackle AVSR and AVS2TT tasks without labeled---transcribed or translated---AV data.

\section{Related Work}

\textbf{Self-supervised audio-only speech representation.} 
Self-supervised learning (\textbf{SSL}) for speech aims to establish a general speech representation with unlabeled speech data applicable to various downstream applications, including speech recognition and spoken language understanding tasks~\cite{Yang2021SUPERBSP}. Many of today's successful SSL approaches~\cite{baevski2020wav2vec,hsu2021hubert,baevski2022data2vec,chiu2022self} are based on masked span prediction. Specifically, wav2vec 2.0~\cite{baevski2020wav2vec} applies a contrastive predictive coding-like~\cite{Oord2018RepresentationLW} loss on masked speech utterances. HuBERT~\cite{hsu2021hubert} utilizes masked speech features to predict hidden units derived from layerwise features. Instead of using discrete units, Data2vec~\cite{baevski2022data2vec} directly regresses dense features from an exponential moving average (EMA) teacher.

SSL approaches notably reduce the need for labeled speech data. In ASR benchmarks~\cite{panayotov2015librispeech}, they yield performance on par with or better than fully supervised models with far fewer transcriptions. This has motivated sustained efforts toward multilingual SSL models such as XLSR-53~\cite{conneau21_interspeech}, XLS-R~\cite{babu22_interspeech}, and MMS~\cite{pratap2023scaling}, which particularly excel in low-resource language scenarios.

\textbf{Self-supervised audio-visual speech representation.} Audio-visual SSL approaches draw heavy inspiration from their audio-only counterparts. AV-HuBERT~\cite{shi2022learning} extends HuBERT~\cite{hsu2021hubert} to the audio-visual setting by taking the masked audio-visual stream as input and predicting the hidden units initialized with MFCC clusters, iteratively refining them with layerwise features. This framework has proven effective for multiple downstream tasks, including lip reading~\cite{shi2022learning}, audio-visual speech recognition and translation~\cite{shi22_interspeech,anwar2023muavic}. In AV2AV \citep{choi2023av2av} and our work of XLAVS-R, it has been extended to a multilingual setting. RAVEn~\cite{haliassos2022jointly} leverages modality-specific EMA teachers to generate targets for masked prediction, thus avoiding an iterative refinement process. Similarly, AV-data2vec~\cite{lian2023av} is based on data2vec~\cite{baevski2022data2vec} and regresses multimodal features with an audio-visual EMA teacher. AV2vec~\cite{zhang2023self} further combines AV-data2vec with the masked prediction objective in AV-HuBERT.
Some prior works also investigated the use of unpaired unimodal data, notably audio, to improve multimodal representation. u-HuBERT~\cite{hsu2022uhubert} augments AV-HuBERT with audio-only speech during pre-training, resulting in improved audio-visual speech recognition performance. VATLM~\cite{zhu2023vatlm} further added text-only data and trained the model on an arbitrary modality stream. 
Our work builds on the insights of AV/u-HuBERT with adaptations for continual SSL training with visual modality injection to A-only model and for the multilingual setting.

\textbf{Audio-visual adaptation of audio-only speech models.} Recent work has begun to explore the adaptation of audio-only speech models into audio-visual models. MixSpeech~\cite{cheng2023mixspeech} develops a visual speech translation model based on a pre-trained audio-only speech translation model by minimizing the discrepancy between the probabilities from audio-only and multi-modal streams. AVFormer~\cite{seo2023avformer} incorporates lightweight modules into an audio-only speech recognizer to adapt it into visually grounded speech recognition through two-stage fine-tuning. FAVA~\cite{may2023audiovisual} directly fine-tunes a pre-trained BEST-RQ~\cite{chiu2022self} encoder for audio-visual speech recognition with a randomly initialized visual encoder. FAVA connects to our work by applying this technique to a multilingual audio-only model. 
However, its evaluation focuses on English with labeled AV speech data for supervised fine-tuning, while our approach focuses on AV SSL to achieve competitive zero-shot results without AV labeled data as well as state-of-the-art supervised multilingual performance.

\begin{figure}[t]
    \centering
    \includegraphics[width=1\columnwidth]{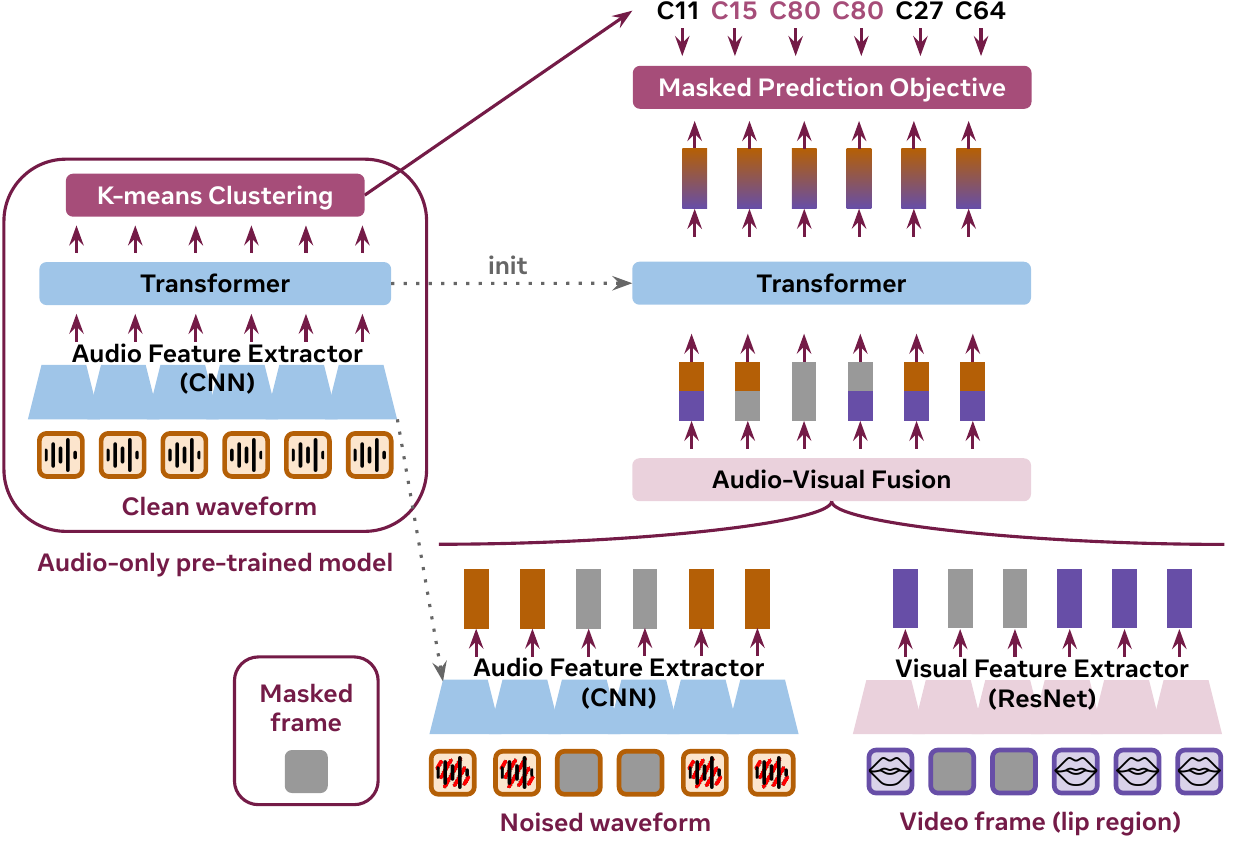}
    \caption{Overview of XLAVS-R. 
    From the audio-only SSL model, we generate unit targets for audio-visual SSL pre-training (left box). We inject visual modality into the first stage model (blue blocks) and fuse visual modality with audio one (pink blocks) to continue training with audio-visual SSL pre-training.
    In AV SSL, noises are added randomly to clean audio, and masked prediction objectives (right top) are applied to the union set of masked frames of audio and visual stream (bottom gray).
    }
    \label{fig:overview}
\end{figure}

\section{XLAVS-R}
\label{sec:xlavsr_method}

Our approach (Figure~\ref{fig:overview}) requires first training an audio-only SSL model, which is used to generate unit targets for audio-visual SSL pre-training. We inject visual modality into the resulting model to continue training with audio-visual SSL pre-training. Through this process, XLAVS-R introduces three key improvements over baseline approaches.
First, we leverage the abundant availability of audio-only speech data to enhance overall performances and achieve greater robustness on domain shift (\textsection \ref{sec:xlavsr_audioonly_speech_rep_learning}), and inject visual modality to build an audio-visual pre-training model on top of audio-only pre-training model (\textsection \ref{sec:xlavsr_visual_modality_injection}).
Second, we use a learnable audio feature extractor to better capture the various phonetic information of multilingual audios (\textsection \ref{sec:xlavsr_Learnable_AFE}).
Third, we improve the training efficiency of the audio-visual pre-training stage by single-round training with unit targets from audio-only contextualized representation (\textsection \ref{sec:xlavsr_single_round_training}).

\subsection{Audio-Only Pre-Training}
\label{sec:xlavsr_audioonly_speech_rep_learning}
Instead of training an audio-visual model in one step with the mix of audio-only data and audio-visual data, we first train an audio-only speech model and adapt it to an audio-visual one via self-supervised continual pre-training with audio-visual data. This is motivated by the facts that audio-only data is much more plentiful than audio-visual data, particularly for low-resource languages, and that audio-only models are much more computationally efficient than audio-visual models with similar architecture and size because of the cost of visual feature extraction and bi-modal feature fusion. We concentrate our computational resources on the first stage of A-only SSL since audio-only data has greater volume and the audio modality naturally contains richer semantic information than the visual modality.
We refer to the outcome model of this audio-only self-supervised learning as XLS-R, as we follow the XLS-R training settings \citep{babu22_interspeech} and adopt its wav2vec 2.0 architecture \citep{baevski2020wav2vec}.

\subsection{Continued Audio-Visual Pre-Training with Visual Modality Injection}
\label{sec:xlavsr_visual_modality_injection}
After obtaining XLS-R through A-only pre-training, we add a ResNet-based visual feature extractor (VFE) and a linear projection-based feature fusion module to build up an audio-visual model and continue AV pre-training. 

For audio-visual self-supervised learning, we follow \citet{shi2022learning,shi22_interspeech} using AV-HuBERT pre-training loss.
Given temporally aligned audio input $I_a \in \mathbb{R}^{T_a \times C_a}$ and video speech input $I_v \in \mathbb{R}^{T_v \times C_v \times W \times H}$, each modality feature extractor encode inputs into same length feature sequences $f_a, f_v \in \mathbb{R}^{T \times D}$. 
Here, audio and visual frames are masked independently by separate masking probabilities.
Frame-wisely fused audio-visual features $Fusion(f_a, f_v) \in \mathbb{R}^{T \times D'}$ are processed by the Transformer \citep{vaswani2017attention} to generate contextualized audio-visual representation.
Target units $z \in \{1, \ldots, K\}^{T}$ for each frame are assigned by an unsupervised clustering method.
Given the output probability $p$ from the fused representation and the target units $z$, the model is trained with the following loss:
\begin{equation}
L = - \sum_{t \in M} \log p_t(z_t) - \alpha \sum_{t \notin M} \log p_t(z_t),
\end{equation}
where $M$ is a union set of masked frames for the audio and visuals stream and $\alpha$ is weighing factor for the unmasked regions.

As in AV/u-HuBERT, we apply modality dropout to encourage the fusion of audio and visual representation space.
In the modality fusion, either sequence of $f_a$ or $f_v$ can be dropped out by the modality dropout probability $p_m$ and filled with zero.
In the case of dropping out one of the modalities, $f_a$ is selected with an audio dropout probability $p_a$ to have $Fusion(0, f_v)$.

With the masked prediction objective on audio-only targets, the continued training of audio-visual SSL aligns audio-visual and visual-only representations to corresponding audio-only representation, which is established in the previous A-only SSL. 

Besides the audio-to-visual representation alignment, clean-to-noisy audio representation is also aligned during the continued AV SSL training phase by using potentially noised audio as audio inputs. This is implemented by randomly adding noises to clean speech audio during the training.

After self-supervised learning, we fine-tune with sequence-to-sequence cross-entropy loss for downstream tasks.

\subsection{Learnable Audio Feature Extractor}
\label{sec:xlavsr_Learnable_AFE}

XLAVS-R has the same model architecture as AV/u-HuBERT except for the nature of the audio features provided as inputs to the Transformer encoder. We jointly train a convolutional audio feature extractor (AFE) as that in wav2vec 2.0~\cite{baevski2020wav2vec} for high-dimensional audio inputs. This differs from AV/u-HuBERT which uses lower-dimensional filterbank features. Our approach provides more capacity for multilingual models and their cross-lingual inferences.

\subsection{Single-Round AV Training with Self-Supervised Audio-Only Targets}
\label{sec:xlavsr_single_round_training}
AV-HuBERT \cite{shi2022learning, shi22_interspeech} and u-HuBERT \citep{hsu2022uhubert} require multiple pre-training rounds, with training targets switching from quantized audio-only local features to quantized audio-visual contextualized representation. 
In each round, the AV model is re-trained from scratch. Hence, the model quality gains across training rounds come from the updated training targets.
While these iterative rounds help the model encode more advanced contextual and bi-modal information in the later stage, the resulting training process is complex and computationally expensive.

To improve training efficiency, we propose to create first-round training targets from a contextualized audio-only representation instead of local features that have noisier, lower-level information. This accelerates the masked prediction learning of high-level semantic information in the first round and reduces the necessity for additional training rounds. We obtain audio-only contextualized representation from a self-supervised multilingual speech model described in Section~\ref{sec:xlavsr_audioonly_speech_rep_learning}. 

\begin{table*}[t]
    \centering
    \fontsize{7.5}{10}\selectfont
    \begin{tabular}{lcrrrrrrrrrrrr}
    \toprule
    \multirow{2}{*}{Model} & \multirow{2}{*}{Test Mode} & OOD & \multicolumn{10}{c}{In-Domain} \\
    \cmidrule(lr){4-13}
    &  & Avg & En & Ar & De & El & Es & Fr & It & Pt & Ru & Avg \\
    \midrule[\heavyrulewidth]
    \multicolumn{13}{c}{\it \textbf{\fontsize{9pt}{9pt} Clean environment, Test WER (\%) $\downarrow$}} \\
    \multirow{2}{*}{AV-HuBERT~\citep{shi22_interspeech}} & A & 41.8 & 2.0 & 89.6 & 53.5 & 47.2 & 18.2 & 20.7 & 21.2 & 22.9 & 46.0 & 35.7\\
    & AV & - & 1.7 & 89.4 & 52.0 & 46.2 & 17.4 & 20.3 & 20.8 & 22.1 & 44.7 & 35.0 \\
    \cmidrule(lr){1-13}
    \multirow{2}{*}{u-HuBERT~\citep{hsu2022uhubert}} & A & 41.4 & 1.9 & 89.3 & 53.3 & 47.2 & 17.9 & 20.7 & 21.6 & 22.5 & 45.8 & 35.6 \\
    & AV & - & 1.9 & 89.3 & 52.1 & 46.4 & 17.3 & 20.5 & 21.2 & 21.9 & 44.4 & 35.0 \\
    \cmidrule(lr){1-13}
    XLS-R 300M & A & 33.2 & 2.5&85.6&44.0&34.4&13.2&15.1&14.3&16.2&34.4&28.8 \\
    \cmidrule(lr){1-13}
    XLAVS-R\\
    \multirow{2}{*}{\quad XLAVS-R 300M} & A & 32.5 & 2.5 & 82.4 & 45.6 & 24.2 & 11.3 & 14.6 & 12.9 & 13.7 & 33.0 & 26.7 \\
    & AV & - & 2.4 & 81.7 & 44.7 & 24.3 & 10.9 & 14.4 & 12.8 & 13.2 & 32.7 & 26.3 \\
    \cmidrule(lr){2-13}
    \cmidrule(lr){2-13}
    \multirow{2}{*}{\quad XLAVS-R 2B} & A & 34.2 & 1.9 & 80.3 & 45.5 & 19.5 & 9.4 & 12.5 & 10.9 & 11.6 & 26.0 & 24.2 \\
    & AV & - & 1.7 & 79.3 & 44.4 & 19.0 & 9.1 & 12.3 & 10.6 & 11.2 & 25.0 & 23.6 \\
    \midrule[\heavyrulewidth]
    
    \multicolumn{13}{c}{\it \textbf{\fontsize{9pt}{9pt}Noise environment, Test WER (\%) $\downarrow$}} \\
    \multirow{2}{*}{AV-HuBERT~\citep{shi22_interspeech}} & A & 89.3 & 39.3 & 111.2 & 87.0 & 85.2 & 65.6 & 58.1 & 69.5 & 69.9 & 75.9 & 73.5 \\
    & AV & - & 6.4 & 104.7 & 74.2 & 70.4 & 43.1 & 43.0 & 48.2 & 47.5 & 67.4 & 56.1 \\
    \cmidrule(lr){1-13}
    \multirow{2}{*}{u-HuBERT~\citep{hsu2022uhubert}} & A & 87.5 & 31.8 & 109.4 & 84.7 & 83.1 & 62.7 & 56.6 & 67.8 & 68.2 & 74.4 & 71.0 \\
    & AV & - & 6.6 & 102.3 & 73.2 & 69.7 & 43.7 & 43.2 & 48.5 & 47.6 & 67.0 & 55.8 \\
    \cmidrule(lr){1-13}
    XLS-R 300M & A & 74.4 & 43.8&97.3&69.8&74.8&47.6&37.1&47.9&54.4&59.8&59.2 \\
    \cmidrule(lr){1-13}
    XLAVS-R\\
    \multirow{2}{*}{\quad XLAVS-R 300M} & A & 79.8 & 42.4 & 100.1 & 69.4 & 56.2 & 39.9 & 33.4 & 40.5 & 45.4 & 53.0 & 53.4 \\
    & AV & - & 5.8 & 95.8 & 61.4 & 44.7 & 28.0 & 27.2 & 29.4 & 30.6 & 47.8 & 41.2 \\
    \cmidrule(lr){2-13}
    \cmidrule(lr){2-13}
    \multirow{2}{*}{\quad XLAVS-R 2B} & A & 74.0 & 49.5 & 98.9 & 66.3 & 50.6 & 36.0 & 30.0 & 36.8 & 40.6 & 48.3 & 50.8 \\
    & AV & - & 5.9 & 93.5 & 58.5 & 38.6 & 23.9 & 23.5 & 24.6 & 26.1 & 41.0 & 37.3 \\
    \bottomrule
    \end{tabular}
    \caption{Results of audio-visual speech recognition (AVSR) on in-domain (MuAViC) and out-of-domain (OOD, FLEURS) evaluation in test mode of audio-only (A) and audio-video (AV).
    Our XLAVS-R model outperforms the two English-only baselines of AV-HuBERT and u-HuBERT (both 300M sized) by up to 18\% WER of in-domain average in noisy AV mode.
    }
    \label{table:sr}
\end{table*}

\section{Experiments}
\label{sec:experimental_setup}
\textbf{Data.}
For AV pre-training, 
we leverage a total 1.2K hours of data in 9 languages: English (En), Arabic (Ar), German (De), Greek (El), Spanish (Es), French (Fr), Italian (It), Portuguese (Pt) and Russian (Ru). 
We also experiment with additional data for domain and language coverage. 
On top of the above 1.2K hour data, we add 7.1K hours of in-house AV data in 100+ languages and train XLAVS-R with a total of 8.3K hours in 100+ languages, as summarized in the last row of Table~\ref{table:data}.
The number of hours we use in the training of each experiment setup for 9 language we focus on are shown in Table~\ref{table:data_stat_appndx}.

For A-only pre-training, we follow 
\citet{babu22_interspeech} 
for training on 436K-hour data in 128 languages.

For all the pre-trained models, we perform multilingual fine-tuning on MuAViC labeled data for audio-visual speech recognition (AVSR) in all 9 languages and audio-visual speech-to-text translation (AVS2TT) in 6 language pairs into English---El, Es, Fr, It, Pt, Ru. 

We use MuAViC for in-domain evaluation on both audio-only and audio-visual test modes.
We use FLEURS~\cite{conneau2023fleurs} for audio-only out-of-domain (OOD) evaluation. We report an average of all 9 languages for AVSR and 6 X-to-English language pairs for AVS2TT in both in-domain and out-of-domain evaluation.

We remove extremely short utterances (less than 0.2 seconds) and long utterances (more than 20 seconds) for better training stability.

\begin{table*}[t]
    \centering
    \fontsize{7.5}{10}\selectfont
    \begin{tabular}{lcrrrrrrrr}
    \toprule
    \multirow{2}{*}{Model} & \multirow{2}{*}{Test Mode} & OOD & \multicolumn{7}{c}{In-Domain}\\
    \cmidrule(lr){4-10}
    & & Avg & El-En & Es-En & Fr-En & It-En & Pt-En & Ru-En & Avg \\
    \midrule[\heavyrulewidth]
    \multicolumn{9}{c}{\it \textbf{\fontsize{9pt}{9pt}Clean environment, Test BLEU $\uparrow$}} \\
    \multirow{2}{*}{AV-HuBERT~\citep{shi22_interspeech}} & A & 12.7 & 13.9 & 22.3 & 28.1 & 23.5 & 26.1 & 10.7 & 20.8 \\
    & AV & - & 14.3 & 22.9 & 28.3 & 23.9 & 26.5 & 11.2 & 21.2 \\
    \cmidrule(lr){1-10}
    \multirow{2}{*}{u-HuBERT~\citep{hsu2022uhubert}} & A & 12.8 & 14.4 & 23.0 & 28.6 & 23.4 & 26.6 & 11.3 & 21.2\\
    & AV & - & 14.5 & 23.1 & 28.6 & 23.7 & 27.0 & 11.4 & 21.4\\
    \cmidrule(lr){1-10}
    XLAVS-R\\
    \multirow{2}{*}{\quad XLAVS-R 300M} & A & 14.0 & 18.9 & 23.7 & 29.7 & 24.9 & 28.5 & 12.6 & 23.0\\
    & AV & - & 19.1 & 23.8 & 29.8 & 25.0 & 28.8 & 13.0 & 23.2 \\
    \cmidrule(lr){2-10}
    \multirow{2}{*}{\quad XLAVS-R 2B} & A & 16.0 & 21.7 & 25.0 & 30.6 & 26.5 & 30.2 & 13.9 & 24.7 \\
    & AV & - & 21.6 & 25.1 & 30.6 & 26.6 & 29.9 & 13.9 & 24.6 \\

    \midrule[\heavyrulewidth]
    \multicolumn{9}{c}{\it \textbf{\fontsize{9pt}{9pt}Noisy environment, Test BLEU $\uparrow$}} \\
    \multirow{2}{*}{AV-HuBERT~\citep{shi22_interspeech}} & A & 3.2 & 4.4 & 9.1 & 13.1 & 8.3 & 8.8 & 4.8 & 8.1 \\
    & AV & - & 8.8 & 15.6 & 19.2 & 15.0 & 17.6 & 7.2 & 13.9 \\
    \cmidrule(lr){1-10}
    \multirow{2}{*}{u-HuBERT~\citep{hsu2022uhubert}} & A & 3.4 & 4.7 & 9.8 & 13.8 & 8.7 & 10.3 & 5.3 & 8.8\\
    & AV & - & 8.2 & 15.6 & 19.7 & 14.6 & 18.3 & 7.3 & 14.0\\
    \cmidrule(lr){1-10}
    XLAVS-R\\
    \multirow{2}{*}{\quad XLAVS-R 300M} & A & 4.3 & 8.3 & 13.9 & 20.4 & 15.2 & 15.1 & 8.2 & 13.5 \\
    & AV & - & 13.2 & 17.4 & 23.8 & 18.7 & 21.8 & 9.4 & 17.4 \\
    \cmidrule(lr){2-10}
    \multirow{2}{*}{\quad XLAVS-R 2B} & A & 6.4 & 11.0 & 15.1 & 20.9 & 16.2 & 16.0 & 9.0 & 14.7\\
    & AV & - & 15.7 & 19.2 & 24.6 & 20.1 & 22.3 & 10.4 & 18.7 \\
    \bottomrule
    \end{tabular}
    \caption{Results of audio-visual speech-to-text translation (AVS2TT) on in-domain (MuAViC) and out-of-domain (OOD, FLEURS) evaluation in test mode of audio-only (A) and audio-video (AV). 
    Our XLAVS-R model outperforms the two English-only baselines of AV-HuBERT and u-HuBERT (both 300M sized) for all language pairs in any mode and by up to 4.8 BLEU of in-domain average in noisy AV mode.
    }
    \label{table:s2tt}
\end{table*}

\paragraph{Modeling.} 
We build XLS-R (\textsection \ref{sec:xlavsr_audioonly_speech_rep_learning}) and XLAVS-R (\textsection \ref{sec:xlavsr_method}) models in two model sizes: 300M and 2B. The number of encoder layers/embedding dimension/feed forward dimension are 24/1024/4096 and 48/1920/7680 respectively. 
For audio-visual training targets of XLAVS-R, we extract audio-only speech representation from the 36th layer of XLS-R 2B and quantize it with 2000 k-means clusters. 
For fine-tuning AVSR and AVS2TT models, we follow \citet{anwar2023muavic} to add a Transformer decoder that has 6 layers, an embedding dimension of 768, and feed forward network dimension of 3072. 
The visual feature extractor of our model is modified ResNet-18 as in AV-HuBERT and prior lipreading works \citep{stafylakis2017combining, martinez2020lipreading, ma2021end}.
For XLAVS-R, we set both $p_m$ and $p_a$ to 0.5 for pre-training which is the same with AV/u-HuBERT. For fine-tuning all pre-trained models, we set $p_m$ and $p_a$ to 0.5 and 0, where the models use 50\% AV modality and 50\% A-only modality. We set $\alpha$ as 1.0 following the AV-HuBERT models\footnote{https://facebookresearch.github.io/av\_hubert}.

\textbf{Noise Augmentation.} 
Following \citet{shi22_interspeech} and \citet{anwar2023muavic}, we randomly augment the input samples with 
additive noises at 0dB of signal-to-noise ratio. Noise audio clips 
are sampled from the MUSAN dataset~\cite{snyder2015musan}
as well as drawn from MuAViC train set for babble type. 
We augment 25\% of the input in pre-training and 50\% in fine-tuning 
with noises.


\textbf{Evaluation.} We evaluate AVSR and AVS2TT in audio-only (``A'') and audio-visual (``AV'') test modes, where the former leverages only audio modality in inference while the latter leverages both audio and visual modalities.
We select the best checkpoint by validation accuracy for inference. We use beam search with default hyper-parameters and beam size of 5. For AVSR, we apply Whisper text normalizer~\cite{radford2022robust} before calculating WER (word error rate). For AVS2TT, we compute BLEU ~\cite{papineni2002bleu} using SacreBLEU~\cite{post-2018-call} with default options and the built-in \textit{13a} tokenizer. For the simulation of noisy test environment, we add babble type of noises drawn from MuAViC test set.

\begin{figure*}[t]
    \centering
    \begin{subfigure}{\linewidth}
        \includegraphics[width=1.0\linewidth]{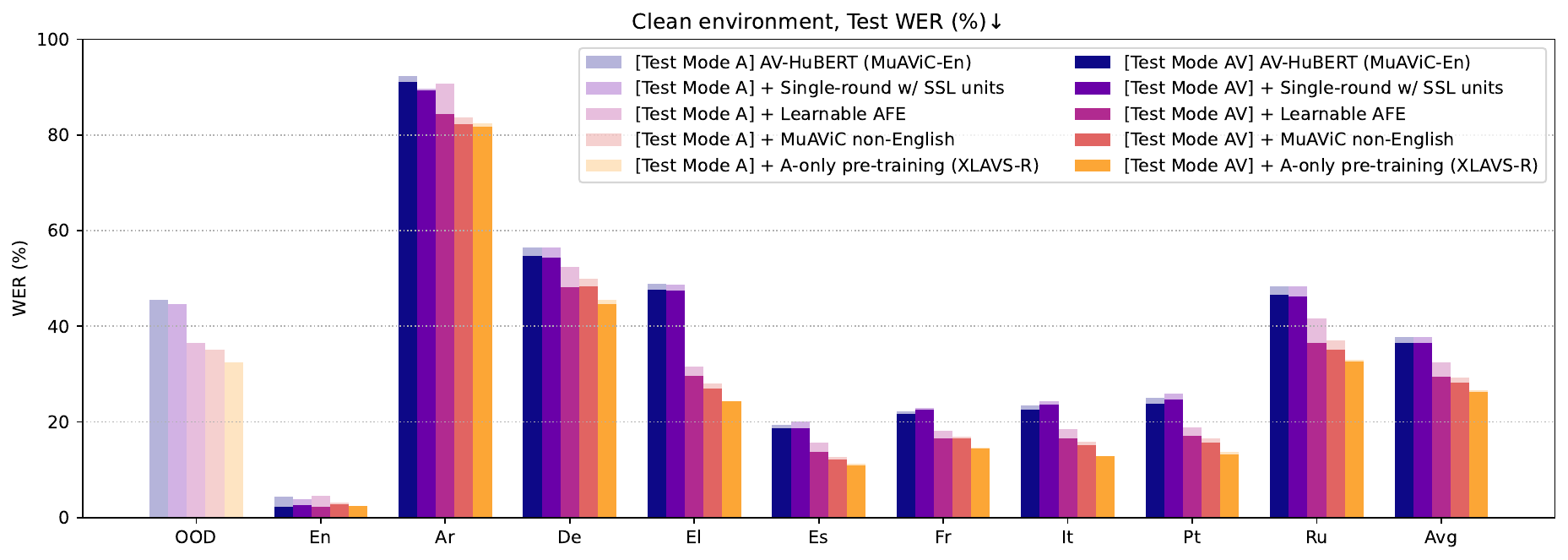} 
    \end{subfigure}
    \begin{subfigure}{\linewidth}
        \includegraphics[width=1.0\linewidth]{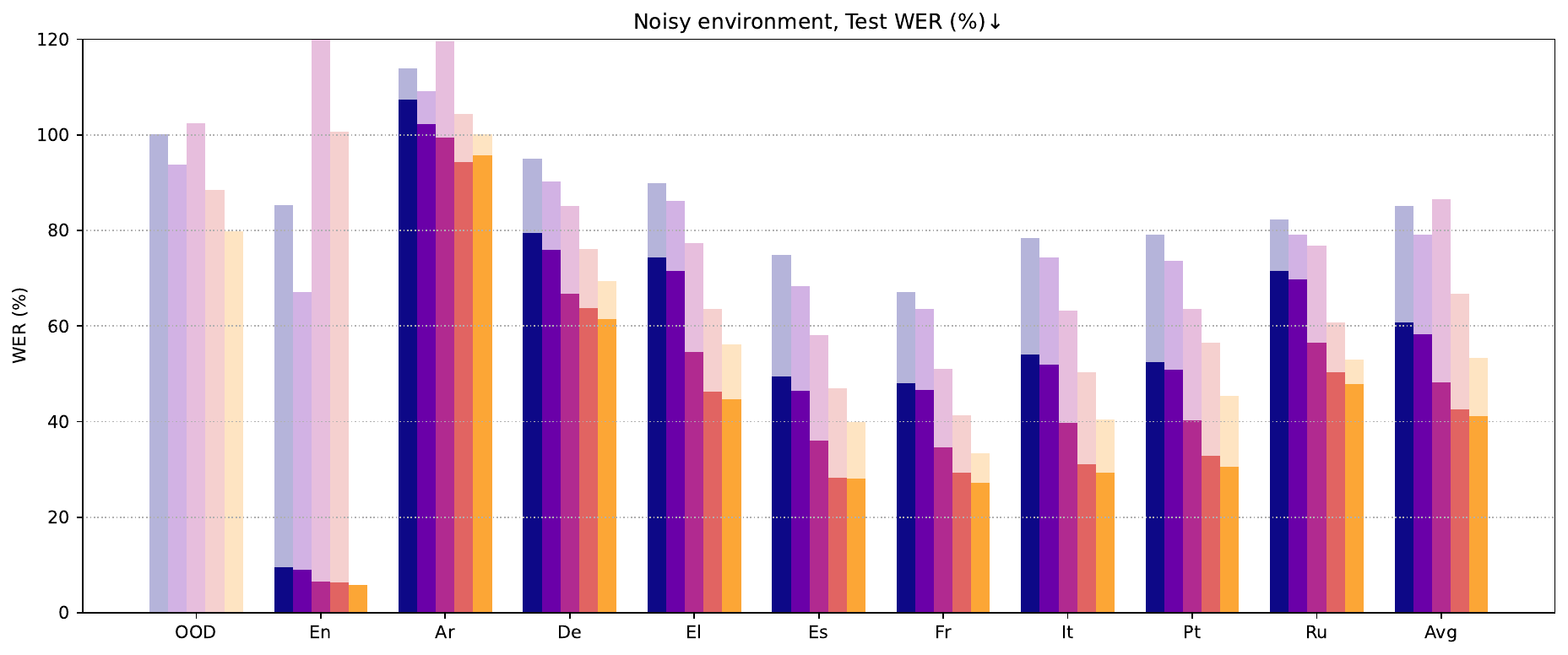} 
    \end{subfigure}
    \caption{Effectiveness of each component towards XLAVS-R and multilingual pre-training data starting from AV-HuBERT
    model pre-trained only on MuAViC-En. 
    All the components of XLAVS-R are shown to be effective.
    Each ablated pre-trained models are fine-tuned and evaluated on multilingual audio-visual speech recognition with identical training and test settings (A: audio, AV: audio+video).
    The numbers of the plots are in Appendix, Table~\ref{table:avhubert2}. 
    }
    \label{fig:avhubert2}
\end{figure*}

\section{Results}

\subsection{Multilingual Speech Recognition}
\label{sec:sr_results}

We compare our XLAVS-R representations against the AV-HuBERT and u-HUBERT English-only representations when fine-tuned for multilingual speech recognition in Table~\ref{table:sr}, reporting WER on both clean and noisy inputs in the upper and lower half of the Table respectively.

In the \textbf{clean setup}, the XLAVS-R models perform on par or better than the baselines in both audio-only and audio-visual test modes. The XLAVS-R 300M model outperforms the AV-HuBERT and u-HuBERT, English-only AV pre-trained models of similar size, by up to 9\% WER and 8.7\% WER respectively for average audio-only and audio-visual test modes. Our larger XLAVS-R 2B outperforms the English-only baseline by an average 11.5\% WER and the average of 11.4\% WER respectively for the two modes. XLAVS-R 300M and 2B both outperform the baselines by large margins in every non-English language. English WER slightly increased in XLAVS-R 300M model compared to the English-only pre-trained baseline by 0.5\% WER. We attribute this effect to inter-language competition \citep{conneau-etal-2020-unsupervised, chang2023multilinguality}, since XLAVS-R 300M covers 128 languages with the same capacity as the English-only baselines (Table~\ref{table:data}). Increasing the XLAVS-R model capacity to 2B achieves the lowest WER of all models in both A and AV modes.
The OOD results on FLEURS confirm that our models also maintain reasonable Audio-only performance on other existing benchmarks.

In the \textbf{noisy setup}, where we add multilingual babble noises to clean speech inputs,  
the XLAVS-R models improve performance across the board, in all languages and in both modes. In the AV mode, the average WER for XLAVS-R 2B drops significantly to 37.3\%, which remarkably approaches that of the AV/u-HuBERT baseline in clean settings.

Overall, XLAVS-R models outperform the two baselines by a large margin on average in clean and noisy settings, while remaining competitive with English-only pre-trained models in English.
Results of experiments with additional AV data for domain and language coverage during AV pre-training are available in Appendix~\ref{sec:additional_training_data} and Table~\ref{table:sr_vc2_avs_appndx}.
We also show that the findings are consistent with different types of noise in Appendix~\ref{sec:noise_type_appdnx}, Table~\ref{table:noise_type_avg_appndx}-\ref{table:noise_type_lang_appndx}.

\begin{figure*}
    \centering
    \subcaptionbox{In-domain test average of 300M sized model\label{fig:xlsr_xlavsr_300m}}%
    [.59\linewidth]{\includegraphics[scale=0.48]{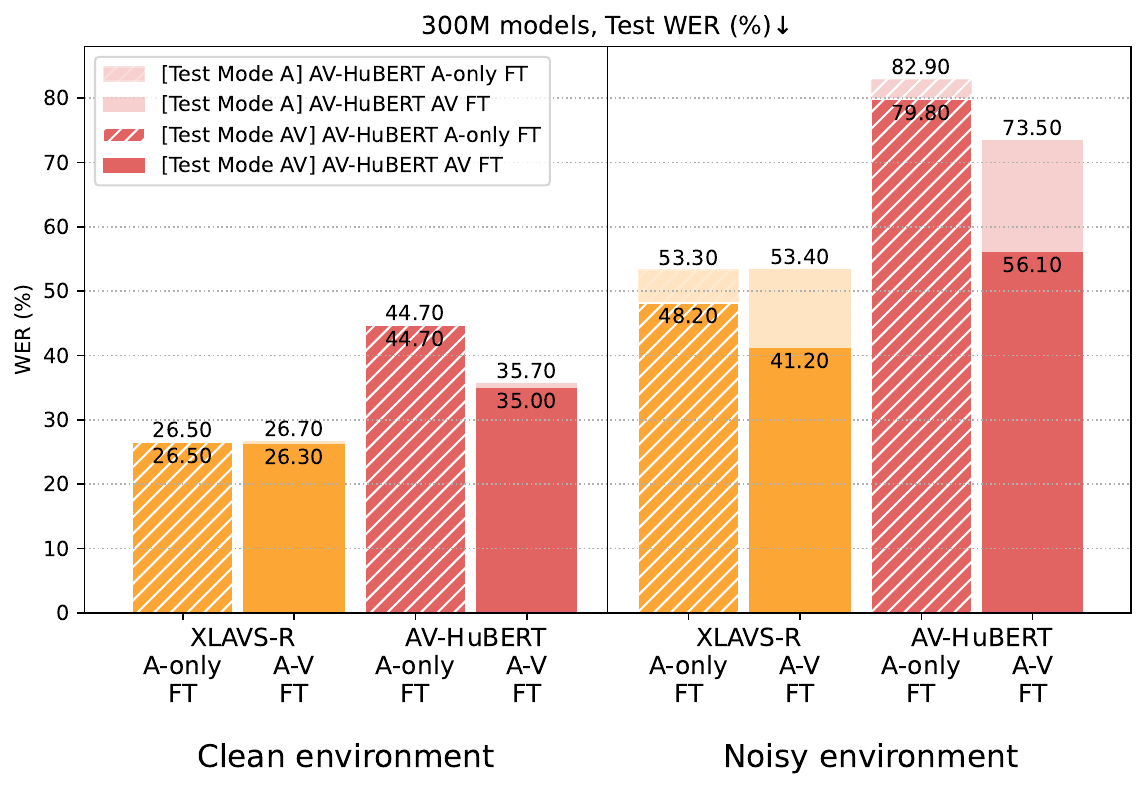}}
    \subcaptionbox{In-domain test average of 2B sized model\label{fig:xlsr_xlavsr_2b}}%
    [.39\linewidth]{\includegraphics[scale=0.48]{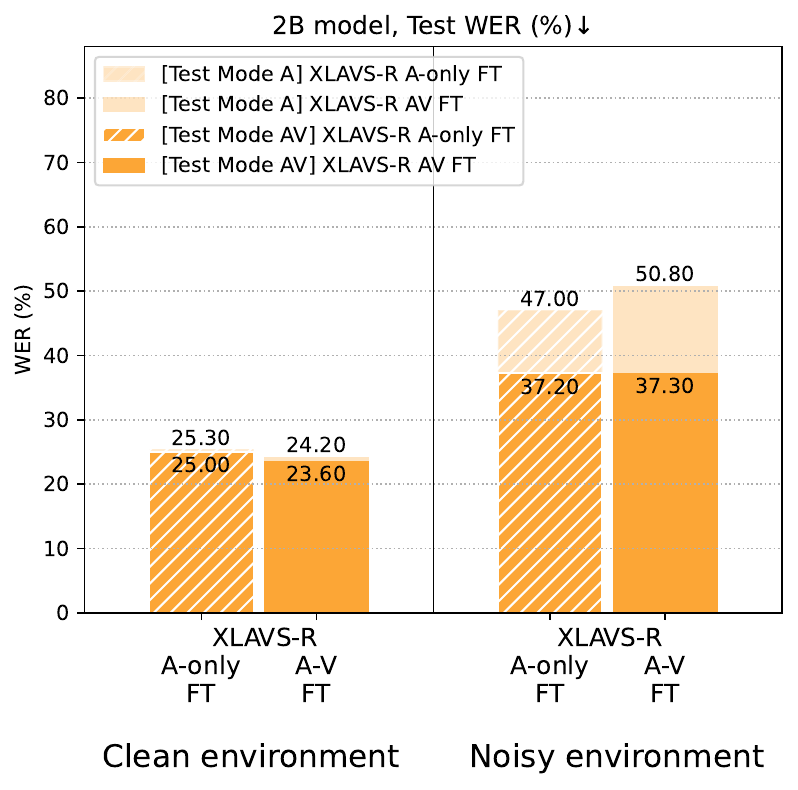}}
    \caption{
    XLAVS-R shows greater zero-shot ability on audio-visual test mode with audio-only fine-tuned (FT, striped) model compared to that of AV-HuBERT. 
    Without fine-tuned on the labeled audio-visual set, the A-only FT model from XLAVS-R shows 5\% WER improvement on AV test mode compared to A test mode (purple), while that of AV-HuBERT shows only 3\% WER (orange) in a noisy environment.
    The bigger the XLAVS-R model size, the greater the zero-shot ability---the gap of 9.8\% WER between A and AV test mode of A-only FT 2B model in a noisy environment).
    Values of individual languages are in Appendix, Table~\ref{table:xlsr_xlavsr}.
    }
    \label{fig:xlsr_xlavsr}
\end{figure*}

\subsection{Multilingual Speech-To-Text Translation}
\label{sec:st_results}

Next, we evaluate models fine-tuned for the X-En AVS2TT tasks (Table~\ref{table:s2tt}), reporting BLEU scores in clean (top) vs. noisy settings (bottom). 

In the \textbf{clean setup}, XLAVS-R 300M outperforms the English-only pre-trained AV/u-HuBERT by up to 2.2 and 2 BLEU average scores for audio-only and audio-visual modes.
It consistently outperforms the baseline by large margins in all directions and test modalities, suggesting that multilingual pre-training is essential for cross-lingual ability in the downstream tasks.  Our best model, XLAVS-R 2B outperforms the English-only baselines by even wider margins (up to 3.9 and 3.4 BLEU average scores respectively for the two modes).

In the \textbf{noisy setup}, we simulate noisy environments as in AVSR tasks by adding multilingual babble noises to clean speech sources.
XLAVS-R 300M outperforms the English-only baselines of similar size largely by up to 5.4/3.5 average BLEU improvement on A/AV, and 6.6/4.8 average BLEU for 2B model.

Overall, these results show that our XLAVS-R models outperform two baselines by a large margin in all translation directions for both clean and noisy settings, both A-only and AV test modes, and both in-domain and OOD.
Results of experiments with additional AV data for domain and language coverage during AV pre-training are available in Appendix~\ref{sec:additional_training_data} and Table~\ref{table:s2tt_vc2_avs_appndx}.

\subsection{Ablation Experiments of XLAVS-R}
\label{sec:ablation}

In Figure~\ref{fig:avhubert2}\footnote{Appendix Table~\ref{table:avhubert2} contains the exact numbers plotted.}, 
we validate the effectiveness of the key changes from AV-HuBERT to XLAVS-R, step-by-step starting from \texttt{AV-HuBERT} pre-trained on \texttt{MuAViC-En}. 
First, simplifying the training process \---\ single-round AV-HuBERT training with self-supervised contextualized audio-only units from XLS-R (\texttt{+ Single-round w/SSL units}) instead of the original 5 training rounds \---\ actually yields moderate improvements.
Second, switching to a learnable audio feature extractor (\texttt{+ Learnable AFE}) shows the biggest improvements in both clean and noisy settings, especially in the low-resource languages (El and Ru).
Third, augmenting AV pre-training with multilingual data (\texttt{+ MuAViC non-English}) is the second-highest improvement indicating that multilingual audio-visual representation learning is critical in the self-supervised pre-training stage.
Last, the introduction of the audio-only pre-training stage and multilingual audio-only resources (\texttt{+ A-only pre-training (XLAVS-R)}) further improves both A and AV performance in the in-domain evaluations. It yields even higher improvements in OOD evaluations, suggesting that audio-only pretraining may enhance robustness against domain shift.

In sum, all of the design contributions of XLAVS-R are shown to be effective.

\subsection{Zero-shot Audio-Visual Inference}
\label{sec:results_zero_shot}

We compare the performance of A-only fine-tuning vs. AV fine-tuning: this matters when labels for a given language in downstream task are not available for AV fine-tuning, and thus we need to rely on the zero-shot audio-visual ability from the pre-trained AV representation.  In A-only FT, we exclude any visual modalities for all languages.

For 300M models (Figure~\ref{fig:xlsr_xlavsr_300m}), XLAVS-R shows only a mild performance degradation in the noisy AV mode from AV FT to zero-shot A-only FT (41.2 WER to 48.2 WER, 17\% increase rate) and almost none in clean. By contrast, the AV-HuBERT baseline shows severe degradation even in the clean setting---56.1 to 79.8, 42.25\% in noisy, and 27.7\% in the clean setting. The XLAVS-R model thus effectively transfers knowledge from the pre-trained AV representation and leverages it for downstream tasks, even though there is no visual modality involved at all during the supervised fine-tuning.

For 2B models (Figure~\ref{fig:xlsr_xlavsr_2b}), although AV FT has better recognition performance in a clean setting (23.6 vs 25.0 WER in AV mode), A-only FT model has even better performances over AV FT mode in noisy settings tested on AV mode (37.2 vs 37.3 WER), showing that the increased model capacity improves its zero-shot ability.
The result of the scaled-up model suggests that XLAVS-R relaxes the need for labeled AV datasets of downstream tasks to have noise robustness.

\begin{table*}[t]
    \centering
    \fontsize{7.5}{10}\selectfont
    \begin{tabular}{lcrrrrrrrrrrrr}
    \toprule
    \multirow{2}{*}{Model} & \multirow{2}{*}{Test Mode} & OOD & \multicolumn{10}{c}{In-Domain} \\
    \cmidrule(lr){4-13}
    &  & Avg & En & Ar & De & El & Es & Fr & It & Pt & Ru & Avg \\
    \midrule[\heavyrulewidth]
    \multicolumn{13}{c}{\it \textbf{\fontsize{9pt}{9pt} Clean environment, Test WER (\%) $\downarrow$}} \\
    XLAVS-R 300M\\
    \quad Basic AV data: 1.2K hours & A & 32.5 & 2.5 & 82.4 & 45.6 & 24.2 & 11.3 & 14.6 & 12.9 & 13.7 & 33.0 & 26.7 \\
    \quad\quad\quad\quad  in 9 languages & AV & - & 2.4 & 81.7 & 44.7 & 24.3 & 10.9 & 14.4 & 12.8 & 13.2 & 32.7 & 26.3 \\
    \cmidrule(lr){2-13}
    \quad Extended AV data: 8.3K hours & A & 28.3 & 2.2 & 80.2 & 38.7 & 28.7 & 11.9 & 15.5 & 14.1 & 14.9 & 32.1 & 26.5 \\
    \quad\quad\quad\quad  in 100+ languages & AV & - & 2.1 & 80.0 & 38.0 & 28.1 & 11.7 & 15.3 & 13.8 & 14.4 & 31.2 & 26.1 \\
    \midrule[\heavyrulewidth]
    
    \multicolumn{13}{c}{\it \textbf{\fontsize{9pt}{9pt}Noise environment, Test WER (\%) $\downarrow$}} \\
    XLAVS-R 300M\\
    \quad Basic AV data: 1.2K hours & A & 79.8 & 42.4 & 100.1 & 69.4 & 56.2 & 39.9 & 33.4 & 40.5 & 45.4 & 53.0 & 53.4 \\
    \quad\quad\quad\quad  in 9 languages & AV & - & 5.8 & 95.8 & 61.4 & 44.7 & 28.0 & 27.2 & 29.4 & 30.6 & 47.8 & 41.2 \\
    \cmidrule(lr){2-13}
    \quad Extended AV data: 8.3K hours & A & 73.7 & 44.0 & 97.2 & 62.4 & 60.7 & 41.0 & 36.2 & 44.5 & 48.0 & 51.0 & 53.9 \\
    \quad\quad\quad\quad  in 100+ languages & AV & - & 5.3  & 91.9 & 53.5 & 49.6 & 28.8 & 29.3 & 32.2 & 32.5 & 46.1 & 41.0 \\
    \bottomrule
    \end{tabular}
    \caption{AVSR comparison of XLAVS-R 300M models trained on different sizes of the AV pre-training dataset. 
    }
    \label{table:sr_vc2_avs_appndx}
\end{table*}

\begin{table*}[t]
    \centering
    \fontsize{7.5}{10}\selectfont
    \begin{tabular}{lcrrrrrrrr}
    \toprule
    \multirow{2}{*}{Model} & \multirow{2}{*}{Test Mode} & OOD & \multicolumn{7}{c}{In-Domain}\\
    \cmidrule(lr){4-10}
    & & Avg & El-En & Es-En & Fr-En & It-En & Pt-En & Ru-En & Avg \\
    \midrule[\heavyrulewidth]
    \multicolumn{9}{c}{\it \textbf{\fontsize{9pt}{9pt}Clean environment, Test BLEU $\uparrow$}} \\
    XLAVS-R 300M\\
    \quad Basic AV data: 1.2K hours & A & 14.0 & 18.9 & 23.7 & 29.7 & 24.9 & 28.5 & 12.6 & 23.0\\
    \quad\quad\quad\quad  in 9 languages & AV & - & 19.1 & 23.8 & 29.8 & 25.0 & 28.8 & 13.0 & 23.2 \\
    \cmidrule(lr){2-10}
    \quad Extended AV data: 8.3K hours & A & 14.4 & 18.5 & 23.6 & 29.6 & 25.1 & 28.6 & 12.0 & 22.9 \\
    \quad\quad\quad\quad  in 100+ languages & AV & - & 18.3 & 23.9 & 29.8 & 25.1 & 28.9 & 12.1 & 23.0 \\

    \midrule[\heavyrulewidth]
    \multicolumn{9}{c}{\it \textbf{\fontsize{9pt}{9pt}Noisy environment, Test BLEU $\uparrow$}} \\
    XLAVS-R 300M\\
    \quad Basic AV data: 1.2K hours & A & 4.3 & 8.3 & 13.9 & 20.4 & 15.2 & 15.1 & 8.2 & 13.5 \\
    \quad\quad\quad\quad  in 9 languages & AV & - & 13.2 & 17.4 & 23.8 & 18.7 & 21.8 & 9.4 & 17.4 \\
    \cmidrule(lr){2-10}
    \quad Extended AV data: 8.3K hours & A & 5.8 & 11.1 & 14.8 & 20.0 & 15.4 & 15.4 & 8.5 & 14.2 \\
    \quad\quad\quad\quad  in 100+ languages & AV & - & 13.8 & 18.4 & 23.7 & 19.6 & 21.7 & 9.8 & 17.8 \\
    \bottomrule
    \end{tabular}
    \caption{AVS2TT comparison of XLAVS-R 300M models trained on different sizes of the AV pre-training dataset. 
    }
    \label{table:s2tt_vc2_avs_appndx}
\end{table*}

\subsection{Results with Additional Training Data}
\label{sec:additional_training_data}
We also experiment with additional data 
for domain and language coverage 
during the audio-visual self-supervised learning phase.

Table~\ref{table:sr_vc2_avs_appndx} shows AVSR performances of 9 languages on OOD and in-domain test sets.
On the in-domain test set, the model pre-trained on a scaled-up AV training set shows improved AV average performances of 0.2\% WER for both clean and noisy environments.
In contrast to the slight improvement or even the degradation for noisy A-only mode in the in-domain, the 8.3K model shows significant improvements in OOD for both clean and noisy settings by up to 6.1\% WER on average.

Table~\ref{table:s2tt_vc2_avs_appndx} shows AVS2TT performances of six X-to-En pairs on OOD and in-domain test sets, where the trend of the prominent improvement in OOD is similar to that of AVSR comparison.
Additional AV SSL dataset improves the average performance of in-domain noisy setting by 0.7 BLEU for A-only test mode and 0.4 BLEU for AV test mode, while slightly degrades in clean settings by 0.1 BLEU in A-only test mode and 0.2 in AV test mode on average.
In out-of-domain evaluation, augmented data in AV SSL enhances the translation quality by up to 1.5 BLEU on average in noisy environments.

Overall, scaling up audio-visual pre-training data enhances robustness against domain shift while also remains effective for noisy robustness with audio-visual input in the in-domain.

\section{Conclusion}

We introduced XLAVS-R, a cross-lingual audio-visual speech representation for noise-robust speech perception in over 100 languages. 
Since audio-visual data is harder to come by than audio-only data, we injected the visual modality by continuing training an audio-only pre-trained model. 
We also design a simpler yet more effective training scheme and improved architecture compared to previous state-of-the-art models. 
Extensive evaluation on the MuAViC benchmark shows XLAVS-R achieves SOTA performance on both audio-visual speech recognition and translation tasks, yielding particularly large improvements in noisy settings.  
We also show that XLAVS-R effectively leverages unlabeled audio-only multilingual speech data and audio-visual data in self-supervised learning resulting in enhanced zero-shot audio-visual ability for downstream tasks, 
and extended AV pre-training data augments robustness against domain shift.

\section*{Limitations}
\label{sec:limitations}

Our findings are inherently limited by the settings of our empirical evaluation. 
For instance, we simulate noisy environments only with the ``babble'' sound in testing experimental setup (\textsection \ref{sec:experimental_setup}), and it remains to be seen how other types of noise might impact our model. In addition, while we include a diverse set of languages with varying amounts of resources in our evaluation, we did not evaluate translation out-of-English and between non-English languages. 

\bibliography{refs}
\clearpage
\appendix

\begin{table*}[t]
\centering
\fontsize{9}{10}\selectfont
\begin{tabular}{crrrrrrrrr}
\toprule
Hours    & En   & Ar  & De  & El & Es  & Fr  & It  & Pt  & Ru  \\
    \midrule[\heavyrulewidth]
Default  & 437  & 19  & 13  & 29 & 181 & 179 & 104 & 156 & 52  \\
Extended & 3378 & 142 & 381 & 40 & 465 & 444 & 228 & 563 & 346 \\\bottomrule
\end{tabular}
    \caption{Statistics of training datasets in hours.}
    \label{table:data_stat_appndx}
\end{table*}

\section{Generalization to Different Types of Noise}
\label{sec:noise_type_appdnx}
As we mentioned in the Limitation section, the evaluation phase in the main tables is conducted on a single type of synthetic noise, Babble.
To confirm the general effectiveness of XLAVS-R on noise robustness, 
we further experiment AVSR with different types of noise, Music and Noise from MUSAN dataset following \citet{shi22_interspeech}. 

Table~\ref{table:noise_type_avg_appndx} and \ref{table:noise_type_lang_appndx} show the results in average and language-wise WER of each models on different types of noise.
Overall, the noise robustness of Audio-Visual speech processing still holds with different kinds of noises. All AV models manage to get a higher quality of speech recognition with visual information than A-only mode in all new types of noise. Babble turns out to be the toughest noisy environment as shown in the above table, while Music and Noise are even able to achieve scores that are close to the clean environment with AV mode, which is consistent with \citet{shi22_interspeech}.
\begin{table*}[]
\centering
\fontsize{8}{10}\selectfont
\begin{tabular}{lcccccccc}
\toprule
Noise Type              & \multicolumn{2}{c}{Clean} & \multicolumn{2}{c}{Babble} & \multicolumn{2}{c}{Music} & \multicolumn{2}{c}{Noise} \\

Mode              & A           & AV          & A            & AV          & A           & AV          & A           & AV          \\
    \midrule[\heavyrulewidth]
AV-HuBERT               & 35.7        & 35.0        & 73.5         & 56.1        & 45.8        & 41.3        & 45.8        & 41.3        \\
u-HuBERT                & 35.6        & 35.0        & 71.0         & 55.8        & 45.1        & 41.0        & 45.4        & 41.0        \\
XLAVS-R 300M            & 26.7        & 26.3        & 53.4         & 41.2        & 33.0        & 30.6        & 33.1        & 30.5        \\
XLAVS-R 300M (extended) & 26.5        & 26.1        & 53.9         & 41.0        & 32.7        & 30.1        & 33.0        & 30.3        \\
XLAVS-R 2B              & 24.2        & 23.6        & 50.8         & 37.3        & 30.4        & 27.8        & 30.5        & 27.7  \\\bottomrule
\end{tabular}
    \caption{Average WER of AVSR with different types of noise of Music and Noise from MUSAN dataset. The noise robustness of XLAVS-R is still the most effective with different kinds of noises.}
    \label{table:noise_type_avg_appndx}
\end{table*}

\begin{table*}[]
\centering
\fontsize{8}{10}\selectfont
\begin{tabular}{llllllllllll}
\toprule
Model                   & Noise Type       & En  & Ar   & De   & El   & Es   & Fr   & It   & Pt   & Ru   & Avg  \\
\midrule[\heavyrulewidth]

\multirow{4}{*}{AV-HuBERT}               & Noisy-A (Music)  & 6.7 & 97.3 & 65.0 & 57.8 & 30.1 & 29.7 & 34.4 & 35.8 & 55.8 & 45.8 \\
                                         & Noisy-AV (Music) & 2.9 & 94.1 & 59.8 & 53.4 & 24.5 & 26.1 & 28.8 & 29.8 & 51.9 & 41.3 \\
                                         & Noisy-A (Noise)  & 7.3 & 96.1 & 65.5 & 58.4 & 29.2 & 29.4 & 34.0 & 36.3 & 56.3 & 45.8 \\
                                         & Noisy-AV (Noise) & 3.2 & 94.3 & 59.7 & 53.6 & 24.2 & 26.5 & 28.1 & 30.0 & 52.4 & 41.3 \\
\midrule
\multirow{4}{*}{u-HuBERT}                & Noisy-A (Music)  & 6.3 & 96.9 & 63.5 & 57.2 & 29.2 & 29.2 & 33.5 & 34.6 & 55.1 & 45.1 \\
                                         & Noisy-AV (Music) & 3.2 & 94.3 & 58.7 & 53.1 & 24.6 & 26.2 & 28.3 & 28.9 & 51.7 & 41.0 \\
                                         & Noisy-A (Noise)  & 6.1 & 96.7 & 64.5 & 58.0 & 28.2 & 30.0 & 33.8 & 35.3 & 56.2 & 45.4 \\
                                         & Noisy-AV (Noise) & 2.9 & 93.5 & 59.1 & 53.9 & 23.8 & 26.6 & 28.1 & 29.2 & 52.0 & 41.0 \\
\midrule
\multirow{4}{*}{XLAVS-R 300M}            & Noisy-A (Music)  & 5.3 & 87.7 & 53.3 & 33.6 & 17.1 & 19.3 & 19.6 & 21.7 & 39.5 & 33.0 \\
                                         & Noisy-AV (Music) & 3.2 & 85.8 & 50.3 & 30.6 & 15.2 & 17.7 & 16.8 & 18.3 & 37.4 & 30.6 \\
                                         & Noisy-A (Noise)  & 6.0 & 87.1 & 54.0 & 33.2 & 17.2 & 19.3 & 19.3 & 21.8 & 40.3 & 33.1 \\
                                         & Noisy-AV (Noise) & 3.5 & 84.8 & 50.7 & 30.0 & 14.8 & 17.9 & 16.4 & 18.1 & 38.0 & 30.5 \\
\midrule
\multirow{4}{*}{XLAVS-R 300M (extended)} & Noisy-A (Music)  & 4.9 & 85.0 & 45.5 & 37.4 & 18.4 & 20.4 & 21.7 & 22.9 & 38.1 & 32.7 \\
                                         & Noisy-AV (Music) & 2.9 & 83.2 & 42.4 & 34.1 & 16.0 & 18.8 & 18.6 & 19.3 & 36.0 & 30.1 \\
                                         & Noisy-A (Noise)  & 5.5 & 85.2 & 46.2 & 37.4 & 18.4 & 20.9 & 21.2 & 23.9 & 38.5 & 33.0 \\
                                         & Noisy-AV (Noise) & 2.9 & 83.1 & 42.8 & 34.2 & 15.7 & 19.3 & 17.9 & 20.0 & 36.5 & 30.3 \\
\midrule
\multirow{4}{*}{XLAVS-R 2B}              & Noisy-A (Music)  & 4.9 & 86.5 & 52.5 & 28.1 & 15.7 & 16.3 & 17.0 & 19.0 & 33.9 & 30.4 \\
                                         & Noisy-AV (Music) & 2.4 & 83.8 & 50.1 & 24.8 & 13.3 & 15.2 & 14.2 & 15.6 & 31.0 & 27.8 \\
                                         & Noisy-A (Noise)  & 5.4 & 86.2 & 52.6 & 27.5 & 15.2 & 16.9 & 17.0 & 19.2 & 34.2 & 30.5 \\
                                         & Noisy-AV (Noise) & 2.5 & 83.0 & 49.3 & 24.9 & 12.7 & 15.2 & 14.2 & 15.6 & 31.9 & 27.7 \\\bottomrule
\end{tabular}
    \caption{Language-wise WER of AVSR with different types of noise of Music and Noise from MUSAN dataset. The noise robustness of XLAVS-R is still the most effective with different kinds of noises.}
    \label{table:noise_type_lang_appndx}
\end{table*}

\begin{table*}[t]
    \centering
    \fontsize{7.5}{10}\selectfont
    \begin{tabular}{lcrrrrrrrrrrrr}
    \toprule
    \multirow{2}{*}{Model} & \multirow{2}{*}{\scriptsize{Test Mode}} & OOD & \multicolumn{10}{c}{In-Domain} \\
    \cmidrule(lr){4-13}
    &  & Avg & En & Ar & De & El & Es & Fr & It & Pt & Ru & Avg \\
    \midrule[\heavyrulewidth]
    \multicolumn{13}{c}{\it \textbf{\fontsize{9pt}{9pt} Clean environment, Test WER (\%) $\downarrow$}} \\
    \multirow{2}{*}{AV-HuBERT (MuAViC-En)} & A & 45.5 & 4.3 & 92.3 & 56.4 & 48.9 & 19.4 & 22.2 & 23.5 & 25.0 & 48.3 & 37.8\\
     & AV & - & 2.2 & 91.1 & 54.7 & 47.7 & 18.6 & 21.6 & 22.6 & 23.8 & 46.5 & 36.5\\
    \cmidrule(lr){2-13}
    \multirow{2}{*}{\quad + Single-round w/ SSL units} & A & 44.6 & 3.9 & 89.7 & 56.4 & 48.7 & 20.1 & 23.0 & 24.4 & 25.9 & 48.3 & 37.8 \\
    & AV & - & 2.6 & 89.3 & 54.4 & 47.5 & 18.7 & 22.5 & 23.7 & 24.6 & 46.3 & 36.6 \\
    \cmidrule(lr){2-13}
    \multirow{2}{*}{\quad + Learned AFE} & A & 36.5 & 4.6 & 90.8 & 52.4 & 31.6 & 15.7 & 18.2 & 18.5 & 18.8 & 41.7 & 32.5 \\
    & AV & - & 2.3 & 84.4 & 48.1 & 29.7 & 13.7 & 16.6 & 16.5 & 17.1 & 36.6 & 29.4 \\
    \cmidrule(lr){2-13}
    \multirow{2}{*}{\quad + MuAViC non-English} & A & 35.1 & 3.1 & 83.7 & 50.0 & 28.1 & 12.7 & 16.9 & 15.9 & 16.6 & 37.1 & 29.3 \\
    & AV & - & 2.8 & 82.2 & 48.4 & 27.0 & 12.1 & 16.6 & 15.2 & 15.6 & 35.1 & 28.3 \\
    \cmidrule(lr){2-13}
    \multirow{2}{*}{\quad + A-only pre-training (XLAVS-R)} & A & 32.5 & 2.5 & 82.4 & 45.6 & 24.2 & 11.3 & 14.6 & 12.9 & 13.7 & 33.0 & 26.7 \\
    & AV & - & 2.4 & 81.7 & 44.7 & 24.3 & 10.9 & 14.4 & 12.8 & 13.2 & 32.7 & 26.3 \\
    \midrule[\heavyrulewidth]
    
    \multicolumn{13}{c}{\it \textbf{\fontsize{9pt}{9pt} Noisy environment, Test WER (\%) $\downarrow$}} \\
    \multirow{2}{*}{AV-HuBERT (MuAViC-En)} & A & 100.2 & 85.3 & 113.9 & 95.0 & 89.9 & 74.8 & 67.1 & 78.5 & 79.2 & 82.3 & 85.1 \\
    & AV & - & 9.5 & 107.3 & 79.5 & 74.3 & 49.4 & 48.0 & 54.1 & 52.5 & 71.5 & 60.7 \\
    \cmidrule(lr){2-13}
    \multirow{2}{*}{\quad + Single-round w/ SSL units} & A & 93.8 & 67.1 & 109.1 & 90.3 & 86.1 & 68.4 & 63.6 & 74.3 & 73.7 & 79.1 & 79.1 \\
    & AV & - & 9.0 & 102.2 & 76.0 & 71.5 & 46.5 & 46.6 & 51.9 & 50.8 & 69.7 & 58.2 \\
    \cmidrule(lr){2-13}
    \multirow{2}{*}{\quad + Learned AFE} & A & 102.4 & 183.6 & 119.5 & 85.2 & 77.4 & 58.1 & 51.1 & 63.2 & 63.5 & 76.9 & 86.5 \\
    & AV & - & 6.5 & 99.5 & 66.8 & 54.6 & 36.1 & 34.7 & 39.7 & 40.2 & 56.6 & 48.3 \\
    \cmidrule(lr){2-13}
    \multirow{2}{*}{\quad + MuAViC non-English} & A & 88.4 & 100.7 & 104.3 & 76.2 & 63.6 & 46.9 & 41.3 & 50.3 & 56.5 & 60.7 & 66.7 \\
    & AV & - & 6.3 & 94.3 & 63.7 & 46.3 & 28.3 & 29.4 & 31.0 & 32.9 & 50.3 & 42.5 \\
    \cmidrule(lr){2-13}
    \multirow{2}{*}{\quad + A-only pre-training (XLAVS-R)} & A & 79.8 & 2.4 & 100.1 & 69.4 & 56.2 & 39.9 & 33.4 & 40.5 & 45.4 & 53.0 & 53.4 \\
    & AV & - & 5.8 & 95.8 & 61.4 & 44.7 & 28.0 & 27.2 & 29.4 & 30.6 & 47.8 & 41.2 \\
    \bottomrule
    \end{tabular}
    \caption{
    This is a table containing all the values of the plots in Section~\ref{sec:ablation}, Figure~\ref{fig:avhubert2}. 
    Effectiveness of each component towards XLAVS-R and multilingual pre-training data starting from AV-HuBERT 
    model pre-trained only on MuAViC-En. 
    All the components of XLAVS-R are shown to be effective.
    Each ablated pre-trained models are fine-tuned and evaluated on multilingual audio-visual speech recognition with identical training and test settings (A: audio, AV: audio+video).
    }
    \label{table:avhubert2}
\end{table*}

\begin{table*}[t]
    \centering
    \fontsize{7.5}{10}\selectfont
    \begin{tabular}{lcrrrrrrrrrrrr}
    \toprule
    \multirow{2}{*}{Model} & \multirow{2}{*}{Test Mode} & OOD & \multicolumn{10}{c}{In-Domain} \\
    \cmidrule(lr){4-13}
    &  & Avg & En & Ar & De & El & Es & Fr & It & Pt & Ru & Avg \\
    \midrule[\heavyrulewidth]
    \multicolumn{13}{c}{\it \textbf{\fontsize{9pt}{9pt} Clean environment, Test WER (\%) $\downarrow$}} \\
    \textit{300M model} \\
    \quad XLAVS-R & A & 30.3 & 2.2 & 81.7 & 45.6 & 23.6 & 11.2 & 14.8 & 12.7 & 13.5 & 32.9 & 26.5 \\
    \quad (A-only fine-tuning) & AV & - & 2.2 & 82.1 & 45.2 & 23.6 & 11.0 & 14.7 & 12.8 & 13.5 & 33.0 & 26.5 \\
    \cmidrule(lr){2-13}
    \quad XLAVS-R & A & 32.5 & 2.5 & 82.4 & 45.6 & 24.2 & 11.3 & 14.6 & 12.9 & 13.7 & 33.0 & 26.7 \\
    \quad (AV fine-tuning) & AV & - & 2.4 & 81.7 & 44.7 & 24.3 & 10.9 & 14.4 & 12.8 & 13.2 & 32.7 & 26.3 \\
    \cmidrule(lr){1-13}
    \textit{2B model} \\
    \quad XLAVS-R & A & 35.5 & 2.2 & 83.3 & 49.0 & 22.3 & 9.6 & 13.2 & 10.5 & 11.2 & 26.6 & 25.3 \\
    \quad (A-only fine-tuning) & AV & - & 2.0 & 83.7 & 48.2 & 21.6 & 9.4 & 13.2 & 10.4 & 10.9 & 26.0 & 25.0 \\
    \cmidrule(lr){2-13}
    \quad XLAVS-R & A & 34.2 & 1.9 & 80.3 & 45.5 & 19.5 & 9.4 & 12.5 & 10.9 & 11.6 & 26.0 & 24.2 \\
    \quad (AV fine-tuning) & AV & - & 1.7 & 79.3 & 44.4 & 19.0 & 9.1 & 12.3 & 10.6 & 11.2 & 25.0 & 23.6 \\
    
    \midrule[\heavyrulewidth]
    
    \multicolumn{13}{c}{\it \textbf{\fontsize{9pt}{9pt} Noisy environment, Test WER (\%) $\downarrow$}} \\
    \textit{300M model} \\
    \quad XLAVS-R & A & 78.1 & 45.1 & 102.2 & 68.1 & 55.1 & 39.2 & 33.2 & 39.9 & 44.9 & 52.3 & 53.3 \\
    \quad (A-only fine-tuning) & AV & - & 13.8 & 101.4 & 66.5 & 52.7 & 37.1 & 31.5 & 38.3 & 40.9 & 51.9 & 48.2 \\
    \cmidrule(lr){2-13}
    \quad XLAVS-R & A & 79.8 & 42.4 & 100.1 & 69.4 & 56.2 & 39.9 & 33.4 & 40.5 & 45.4 & 53.0 & 53.4 \\
    \quad (AV fine-tuning) & AV & - & 5.8 & 95.8 & 61.4 & 44.7 & 28.0 & 27.2 & 29.4 & 30.6 & 47.8 & 41.2 \\
    \cmidrule(lr){1-13}
    \textit{2B model} \\
    \quad XLAVS-R & A & 74.3 & 34.2 & 100.9 & 65.1 & 52.6 & 31.8 & 26.7 & 31.7 & 35.4 & 44.8 & 47.0 \\
    \quad (A-only fine-tuning) & AV & - & 4.3 & 96.3 & 60.3 & 41.6 & 23.0 & 22.1 & 22.9 & 24.7 & 39.3 & 37.2 \\
    \cmidrule(lr){2-13}
    \quad XLAVS-R & A & 74.0 & 49.5 & 98.9 & 66.3 & 50.6 & 36.0 & 30.0 & 36.8 & 40.6 & 48.3 & 50.8 \\
    \quad (AV fine-tuning) & AV & - & 5.9 & 93.5 & 58.5 & 38.6 & 23.9 & 23.5 & 24.6 & 26.1 & 41.0 & 37.3 \\
    \bottomrule
    \end{tabular}
    \caption{
    This is a table containing all the values of the summarized plots in Section~\ref{sec:results_zero_shot}, Figure~\ref{fig:xlsr_xlavsr}.
    XLAVS-R shows greater zero-shot ability on audio-visual test mode with audio-only fine-tuned (FT) model compared to that of AV-HuBERT. 
    Without fine-tuned on the labeled audio-visual set, the A-only FT model from XLAVS-R shows 5\% WER improvement on AV test mode compared to A test mode (purple), while that of AV-HuBERT show only 3\% WER (orange) in noisy environment.
    The bigger the XLAVS-R model size, the greater the zero-shot ability---the gap of 9.8\% WER between A and AV test mode of A-only FT 2B model in a noisy environment).
    }
    \label{table:xlsr_xlavsr}
\end{table*}

\section{Future Works: Visual-only Experiment}
The main focus of this research is on having the best noise-robust speech recognition and translation performance by leveraging stronger audio components and visual information, rather than visual-only input. 
Therefore, the presented AV models may not provide significant visual-only speech recognition capabilities, especially in non-English languages. 
The results of visual-only experiment in English are 52.4 WER for AV-HuBERT and 55.9 WER for u-HuBERT, and non-English results are around 90-100 WER.
We plan to conduct comprehensive VSR experiment with XLAVS-R and further explore how to develop zero-shot visual speech recognition (VSR) ability while maintaining state-of-the-art audio-visual performances as a future work.

\end{document}